\documentclass[10pt, aps, prd, superscriptaddress, nofootinbib,
               amsmath, amsmath,amssymb,
               twocolumn,
               preprintnumbers,
               showpacs,
floatfix,
]{revtex4-2}

\providecommand{\exclude}[1]{}
\usepackage{tabularx,colortbl}
\usepackage{etoolbox}
\usepackage{suffix}   
\usepackage{pgf}  

\newtoggle{mypaper}
\toggletrue{mypaper}

\iftoggle{mypaper}{
    \usepackage[aps, arXiv, todo, secnoindent,
    ]{my_paper}
  \iftoggle{arXiv}{
  }{
    
  }
}{
  \usepackage{graphicx}
  \usepackage{booktabs}
  \usepackage{xcolor, colortbl}
  \definecolor{Gray}{gray}{0.9}

  \usepackage{siunitx}
  \sisetup{
    round-mode = none,   
    mode = text,         
    detect-all = true,            
    detect-mode = false,          
    list-units = single,  
    unit-mode = text,
    number-mode = math,
    range-units = repeat,
  }

  \usepackage[nameinlink]{cleveref}
  
  \providecommand{\mycaptionhook}{}
  \let\oldcaption\caption
  \renewcommand{\caption}[2][]{\oldcaption[#1]{\mycaptionhook{}#2}}

  \usepackage[version=3]{mhchem}

  \providecommand{\mysc}[1]{\MakeUppercase{#1}}
  \providecommand{\upsc}[1]{\MakeUppercase{#1}}

}

\usepackage{graphicx}
\graphicspath{{}{images/}}

\exclude{
  \usepackage{dcolumn}
  \usepackage{bm}
  \usepackage{multirow}

  \usepackage{soul}

}

\usepackage{my_acronyms}
\usepackage[utf8]{inputenc}

\renewcommand{\d}{\mathrm{d}}

\newcommand{\mat}[1]{\bm{#1}}

\newcommand{\dEFT}{{\gls{chiEFT}}}
\newcommand{\Eres}{{\mathfrak{E}}}
\newcommand{\dQMC}{{phenomenological}}
\renewcommand{\dQMC}{{$\phi$\gls{QMC}}}
\newcommand{\FRZ}[2]{\ensuremath{\smash{\text{FRZ}^{#1}_{#2}}}}

\AtEndPreamble{
  \crefname{appendix}{Appendix}{Appendices}
  \Crefname{appendix}{Appendix}{Appendices}
}
\begin{document}
\preprint{LIGO-P2300061}
\preprint{INT-PUB-23-012}

\title{Framework for Multi-messenger Inference from Neutron Stars:\\
  Combining Nuclear Theory Priors}

\author{Praveer Tiwari}
\email{praveer.tiwari@iitb.ac.in}
\affiliation{Department of Physics, Indian Institute of Technology Bombay, Mumbai, Maharashtra, India 400076}
\affiliation{Department of Physics and Astronomy, Washington State University, Pullman, WA, USA 99164}

\author{Dake Zhou}
\affiliation{Department of Physics, University of California, Berkeley, CA, USA 94720}
\email{dkzhou@berkeley.edu}

\author{Bhaskar Biswas}
\affiliation{Inter-University Centre for Astronomy and Astrophysics, Post Bag 4, Ganeshkhind, Pune 411 007, India}
\affiliation{The Oskar Klein Centre, Department of Astronomy, Stockholm University, AlbaNova, SE-10691 Stockholm,
Sweden}
\affiliation{Universit\"{a}t Hamburg, D-22761 Hamburg, Germany}
\email{phybhaskar95@gmail.com}

\author{Michael McNeil Forbes}
\affiliation{Department of Physics and Astronomy, Washington State University, Pullman, WA, USA 99164}
\affiliation{Department of Physics, University of Washington, Seattle, WA, USA 98105}
\email{m.forbes@wsu.edu}

\author{Sukanta Bose}
\affiliation{Department of Physics and Astronomy, Washington State University, Pullman, WA, USA 99164}
\affiliation{Inter-University Centre for Astronomy and Astrophysics, Post Bag 4, Ganeshkhind, Pune 411 007, India}
\email{sukanta@wsu.edu}

\date{\today}

\glsunset{NICER}
\begin{abstract}

We construct an efficient parameterization of the pure neutron-matter \gls{EoS} that incorporates the uncertainties from both \gls{chiEFT}
and phenomenological potential
calculations. 
This parameterization yields a family of \glspl{EoS} including and extending the forms based purely on these two 
calculations.
In combination with an agnostic inner core \gls{EoS}, this parameterization is used in a Bayesian inference pipeline to obtain constraints on the \gls{EoS} parameters
using multi-messenger observations of neutron stars.
We specifically considered observations of the massive pulsar J0740+6620, the binary neutron star coalescence
\gls{GW170817}, and the calculations on the \gls{NICER} data from
pulsar
\textsc{psr j0030+0451} and \textsc{psr j0740+6620}.
Constraints on neutron star mass-radius relations are obtained and compared.
The Bayes factors 
for the different \gls{EoS} models are also computed. 
While current constraints do not reveal any significant preference among these models, the framework developed here may enable future observations with more sensitive detectors to discriminate them.
\end{abstract}

\maketitle

\glsunset{LIGO}
\glsunset{NICER}
\glsreset{EoS}
\section{Introduction}
Astrophysical observations of gravitational waves from the binary neutron star merger \gls{GW170817}~\cite{TheLIGOScientific:2017qsa,abbott2019properties} by the \gls{LIGO}-Virgo collaboration~\cite{LIGOScientific:2014pky,VIRGO:2014yos}, its electromagnetic counterparts~\cite{Abbott_2017, abbott2020model}, and millisecond-pulsar X-rays by \gls{NICER}~\cite{Riley:2019yda, Miller:2019cac, riley2021nicer, Miller2021}, provide multi-messenger information about the structure of neutron stars.
These observations, combined with advances in nuclear theory, give a fillip to the quest of characterizing neutron-rich matter.
This quest requires a framework to calculate observables like the neutron star mass $M$, radius $R$, and tidal deformability $\Lambda$, from a nuclear \gls{EoS} that can incorporate priors from different forms of nuclear theory.
Here we consider priors from \gls{QMC} with phenomenological potentials, and chiral effective field theory \gls{chiEFT}.

Experiments and nuclear theory efforts have constrained the nuclear \gls{EoS} at lower densities, but leave considerable uncertainties above the nuclear saturation density $n_0\approx\qty{0.16}{fm^{-3}}$.
Phenomenological models based on two- and three-body nucleon forces~\cite{PhysRevC.51.38, PhysRevLett.74.4396, PhysRevC.103.065804, PhysRevD.106.083010} have traditionally been used to extrapolate to high density ($\sim5n_0$~\cite{PhysRevD.106.083010}) with somewhat {\it ad hoc} uncertainty quantification.
More recently, \gls{chiEFT} has demonstrated rather accurate predictions with quantified uncertainties for low-density \gls{EoS}~\cite{Drischler:2020}, but questions of convergence cause these uncertainties
to grow rapidly at higher densities: These are currently worse than \num{25}\% above $2 n_0$~\cite{doi:10.1146/annurev-nucl-102419-124827, doi:10.1146/annurev-nucl-102419-041903, PhysRevC.102.055803}.
This has led to a proliferation of neutron star \gls{EoS} models used in the astrophysics
community~\cite{Read:2009}, many of which are poorly linked to the underlying physics.

Recent works assume either specific nuclear models or agnostic functionals to constrain the \gls{EoS} of neutron stars with astrophysical observations.  (See e.g.~\cite{Read:2009, huth2022constraining, dietrich2020multimessenger, pang2021nuclear, pang2022nmma, biswas2021towards, ghosh2022multi, Raaijmakers:2020, huth2022constraining,kat2020, biswas2021towards, biswas2022bayesian, ghosh2022multi, dietrich2020multimessenger, pang2022nmma, pang2021nuclear, landry2020nonparametric, raaijmakers2021constraints, legred2021impact,biswas2021impact}
.)
These typically depend on a dominant parabolic expansion centered around symmetric nuclear matter to constrain the neutron-matter \gls{EoS}.
Here, we instead extend the \gls{FRZ} \gls{EoS} model of~\cite{Forbes:2019} that constructs a physically motivated \gls{EoS} by expanding about pure neutron matter, incorporating the properties of symmetric nuclear matter as constraints on this expansion.
This provides an alternative perspective that focuses on neutron-rich matter, including only the minimal dependence required to be consistent with symmetric nuclear matter.
Our extension adds sufficient flexibility to incorporate both \gls{chiEFT}~\cite{Drischler:2020} results (which we hereafter refer to as \dEFT{} data) and phenomenological potential-based \gls{QMC} calculations (which we hereafter refer to as \dQMC{} data using $\phi$ as a mnemonic for phenomenological)~\cite{Gandolfi:2009, Gandolfi:2010, Gandolfi:2012}).
Our model smoothly interpolates between these data with a parameter $\zeta \in [0, 1]$.
This allows us to directly infer the underlying nuclear \gls{EoS} in the Bayesian framework with physically motivated priors.
We additionally extend the \gls{FRZ} \gls{EoS} with a more generalized \gls{EoS} for the inner core $n>n_c$
consisting of 3 piecewise polytropes that we smoothly attach  
at a transition density chosen from
$n_c \in \{1.5n_0, 2.0n_0\}$.

\paragraph*{Nomenclature} We refer to the last \gls{EoS} described above as \FRZ{\zeta}{n_c/n_0}, where $\zeta=0$ directly models the \dEFT{} data~\cite{Drischler:2020}, $\zeta=1$ directly models the \dQMC{} data~\cite{Gandolfi:2009, Gandolfi:2010, Gandolfi:2012}, and $n_c/n_0 \in \{1.5, 2.0\}$ describes the density $n_c$ where the transition to the polytropic inner-core occurs.
Thus, we refer to models like $\FRZ{\chi}{1.5} \equiv \FRZ{\zeta=0}{1.5}$ 
to denote those
that are fit to the \dEFT{} data with a core-transition at $n_c=1.5n_0$, and $\FRZ{\phi}{2.0} \equiv \FRZ{\zeta=1}{2.0}$ for models fit to the \dQMC{} data with a core transition at $n_c = 2.0n_0$, etc.
We refer to models where we let $\zeta$ vary as ``hybrid'' models.


\section{Equation of state}\label{sec:eos}
The composition of a neutron star varies as the pressure increases with depth.
In the outer crust, the mass and neutron number of nuclei increase with pressure.
Crossing to the inner crust at the drip line~\cite{Thoennessen:2004rpp}, neutrons drip out of the nuclei and accumulate in the interstitial space~\cite{hansel2007neutron}, forming a neutron superfluid that is well modeled as a \gls{UFG}~\cite{Zwerger:2012}.
Deeper nuclei deform, forming regions of nuclear ``pasta''~\cite{Caplan:2017} that eventually merge into a homogeneous phase of protons, neutrons, and electrons at the start of the outer core.
In the inner core when $n \gtrsim 2n_0$, the physics becomes uncertain:
Plausible speculations include transitions to hyperonic matter, strange quark matter,
and color superconducting matter~\cite{Page:2006, Fukushima:2011, PhysRevD.81.105021, lattimer2021neutron}.

To describe the interior of neutron stars, efforts have been made to build unified effective \gls{EoS}s~\cite{Fortin:2016, Zdunik:2016, Forbes:2019} that self-consistently bridge the tabulated properties of the outer crust~\cite{1971ApJ...170..299B} with the homogeneous outer core by using a \gls{CLDM}. 
The \gls{CLDM}
models the transition from nuclei to homogeneous matter with a spherical Wigner-Seitz cell containing the nuclei, the interstitial neutron superfluid, and the lepton gas.
While this precludes a formal analysis of more complicated pasta phases and related dynamical phenomena like shear modes associated with these rigid structures, it provides a quantitatively accurate characterization of the \gls{EoS} because the energy difference between competing pasta phases is small~\cite{Caplan:2017, fattoyev2017quantum}.

As mentioned above, a unique property of the unified \gls{FRZ} \gls{EoS}~\cite{Forbes:2019} that we extend here is the expansion about pure neutron matter.
As shown in~\cite{Forbes:2019} and confirmed by our global sensitivity analysis in \cref{sec:glob-sens-analys}, this approach is quite insensitive to the properties of symmetric nuclear matter, allowing astrophysical observables to directly constrain the properties of neutron-rich matter.
While consistent with global fits to nuclear data~\cite{Bulgac:2018}, this insensitivity differs markedly from analyses that assume a dominant quadratic iso-spin expansion.
One appeal of a dominant quadratic expansion is that one can directly relate the full symmetry energy $S$ and its slope $L$ with the corresponding coefficients $S_2$ and $L_2$ in the quadratic expansion about the symmetric nuclear matter that is constrained with terrestrial experiments.

While this quadratic dominance has some support at low densities~\cite{Wlazlowski:2014a} and appears to hold in \gls{EFT}-based calculations~\cite{Drischler:2020}, recent tension arising from neutron skin measurements~\cite{PREX:2021umo} highlight the subtleties and challenges in faithfully extracting $S$ and $L$.
Our approach is complementary, directly connecting observations with the full parameters $S$ and $L$, but breaking the connection with $S_2$ and $L_2$.
Further explicit assumptions about the iso-spin expansion must be added to our model if any significant connection between observations and $S_2$ and $L_2$ are desired~\cite{Forbes:2019}.

We extend the \gls{FRZ} \gls{EoS} in two ways.
First, we find a new parameterization of the pure neutron-matter \gls{EoS} that consistently incorporates both the original double polytropic parameterization of~\cite{Gandolfi:2012}, and state-of-the-art \gls{EFT} calculation at \gls{N3LO}~\cite{Drischler:2020} with a single new parameter $\zeta$ 
that 
smoothly interpolates from one dataset to the other.
Second, we update the inner core model, replacing the quadratic speed-of-sound \gls{EoS} that was artificially constrained by a conjectured analogy with finite temperature \gls{QCD}, with a piecewise polytrope parameterization~\cite{Raaijmakers:2020, Read:2009}  that allows for greater variations in this poorly understood region.

\subsection{Pure Neutron Matter}\label{sec:eosPNM}
We characterize the \gls{EoS} of pure neutron matter in terms of the energy per particle $E_N = \mathcal{E}(n_n)/n_n$, where $\mathcal{E}$ is the energy density and $n_n$ the neutron number density.
While the energy calculated (\dQMC{}~\cite{Gandolfi:2012, unpublishedGandolfi}) from the phenomenological potentials ~\cite{Gandolfi:2009, Gandolfi:2010, Gandolfi:2012} can be modeled well by the double polytrope $E_N \propto a n_n^\alpha + b n_n^\beta$ used in~\cite{Forbes:2019}, this form is not sufficiently flexible to accurately model the \gls{chiEFT} data~\cite{Drischler:2020}.
We found a polynomial fit to the energy per neutron that captures both the central values and correlated uncertainties of both models.

\begin{figure}
    \centering
    \includegraphics{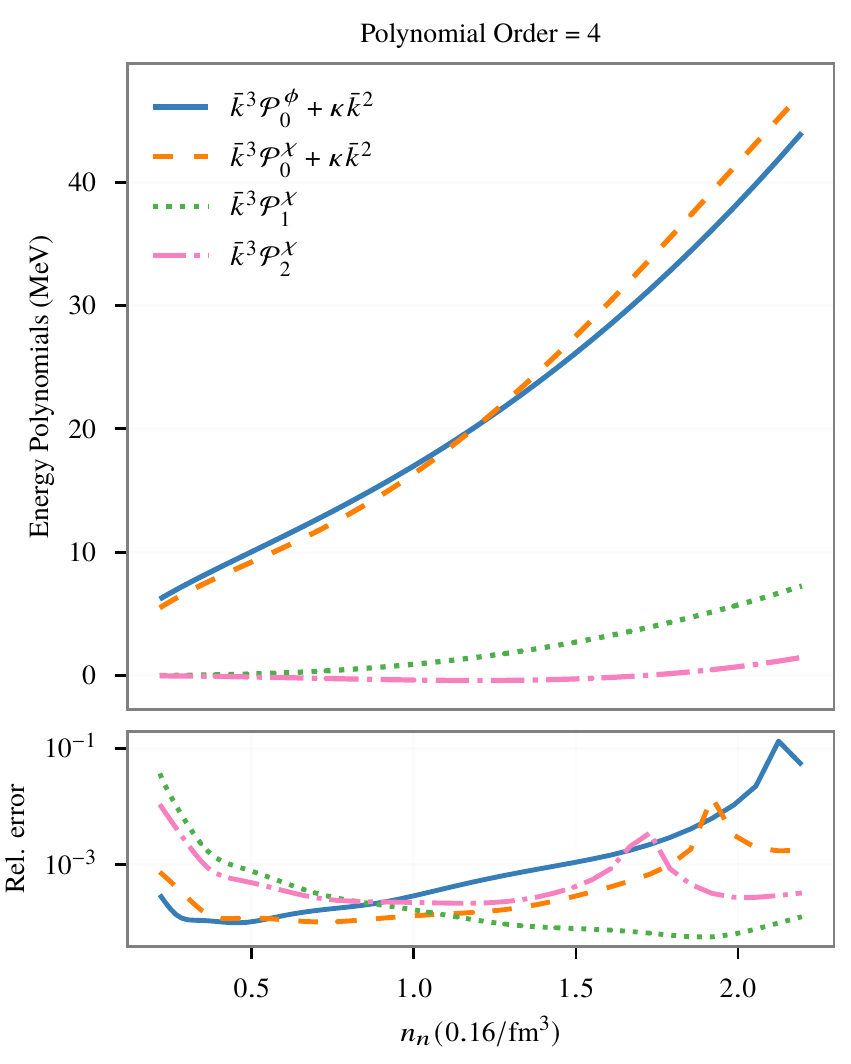}
    \caption{
      The fourth-order polynomial 
      $\Eres(\bar{k})$,
which defines the energy per neutron in Eq.~(\ref{Eq:EPNM}),
      is expanded in the basis of $\Eres_{1, 2}$ (see Eq.~(\ref{eq:Polys})) about the mean $\Eres_0$. The basis functions $\Eres_{1,2}$ are fourth-order polynomials in $\bar{k}$ (plotted with dotted and dash-dot lines) obtained by fitting the two most significant vectors of the covariance matrix of $(E_N (\bar{k}) - m_{n} c^2 -\kappa \bar{k}^2)/\bar{k}^3$ coming from \dEFT{} data. The means for the \dEFT{} and \dQMC{} data are  fitted with fourth-order polynomials (in $\bar{k}$) to obtain $\Eres_{0}^{\chi}$ (dashed line) and $\Eres_{0}^{\phi}$ (solid line). Also plotted (thin solid lines near the dotted line) are the deviations from $\Eres_{0}^{\phi}$ for different realizations of \dQMC{} \glspl{EoS}.
}
    \label{fig:PNM_Fits}
\end{figure}
We model the energy per neutron as
\begin{gather}
  E_N(\bar{k}) = m_{n}c^2 + \kappa \bar{k}^2 + \bar{k}^3 \Eres(\bar{k}),
  \label{Eq:EPNM}
\end{gather}
where $\Eres(\bar{k})$ is a fourth-order polynomial of the dimensionless parameter $\bar{k} \equiv (n_n/n_0)^{1/3}$, which is proportional to the Fermi momentum.
The first $\kappa \bar{k}^2$ term ensures that \cref{Eq:EPNM} approaches the \gls{EoS} of the \gls{UFG} at low densities.
This is consistent with the analysis in~\cite{Bulgac:2018} that we use to fix the value of $\kappa = \SI{25.5}{MeV}$.

For the \dEFT{} data, \textcite{Drischler:2020} characterize the truncation errors with a Gaussian process~\cite{mackay2003information}.
This gives a covariance matrix with two dominant principle components that account for
more than \num{95}\% of the uncertainties, which is sufficient for current astrophysical
analyses.
To this level, we can thus faithfully characterize this \dEFT{} dataset with three fixed
polynomials (see \cref{fig:PNM_Fits}):
\begin{equation}
  \Eres^{\chi}(\bar{k}) = \Eres^{\chi}_0 (\bar{k}) + a_1\Eres_1(\bar{k}) + a_2\Eres_2(\bar{k}),
  \label{eq:Polys}
\end{equation}
where $\Eres^{\chi}_0 (\bar{k})$ fits the central values, and $\Eres_{1,2}(\bar{k})$ fit the two dominant principal components of the covariance matrix.
We find that fourth-order polynomials are sufficient, and scale $\Eres_{1,2}(\bar{k})$ appropriately so that $a_{1, 2} \in [-2, 2]$ captures the $2\sigma$ uncertainty region.

For the \dQMC{} data, \textcite{Gandolfi:2009, Gandolfi:2010, Gandolfi:2012} performed multiple \gls{QMC} calculations for a variety of phenomenological potentials that include both two- and three-body contributions.
As we discuss in \cref{sec:QMCData}, all of these data can also be efficiently
described by three analogous fourth-order polynomials: $\smash{\Eres^{\phi}_0} (\bar{k})$ fitting
the central values, and $\smash{\Eres^{\phi}_{1,2}}(\bar{k})$ fitting the correlated variations
in the data.

Surprisingly, although the central values $\smash{\Eres^{\phi}_0} (\bar{k})$ differ from $\smash{\Eres^{\chi}_0} (\bar{k})$, it turns out that the spread in the \dQMC{} data is well described by the \emph{same} uncertainty model $a_{1,2} \in [-2, +2]$ with the same $\Eres_{1,2}(\bar{k})$ as the \dEFT{} data, (though not with the same rigorous meaning of $2\sigma$).
Perhaps this surprising relationship hints at a consistency between the two physically-inspired microscopic approaches, or provides insight into the physics of many-nucleon systems.

In any case, we can thus accurately parameterize both sets of data with a three-parameter model $(\zeta, a_1, a_2)$:
\begin{equation}   
  \label{eq:PolysQNM}
  \Eres(\bar{k}) = \Bigl(\zeta\Eres_0^{\phi}(\bar{k}) + (1-\zeta)\Eres_0^{\chi}(\bar{k})\Bigr)
  + a_1 \Eres_1(\bar{k}) + a_2 \Eres_2(\bar{k})
\end{equation}
where the parameter $\zeta \in [0, 1]$ smoothly interpolates between \dQMC{} ($\zeta=0$)
and \dEFT{} ($\zeta=1$).  The fourth-order polynomials $\smash{\Eres_0^{\phi}}(\bar{k})$,
$\smash{\Eres_0^{\chi}}(\bar{k})$, $\Eres_1(\bar{k})$, and
$\Eres_2(\bar{k})$ and the fitting process are described in
\cref{sec:ChiData,sec:QMCData}.

\subsection{Inner Core}\label{sec:eoscore}\noindent%
At baryon number densities $n > n_c$ we use a set of three piecewise polytropes
$P_i(n)$~\cite{Hebeler:2013nza, Raaijmakers:2020} to describe homogeneous matter in
beta-equilibrium:
\begin{subequations}
  \begin{gather}
    \nonumber\\[-0.5\baselineskip]
    P(n) = \begin{cases}
      P_1(n) & \!\!\!\!\smash{\overbrace{n_1}^{n_1\equiv n_c}}\!\!\!\! < n \leq n_2,\\
      P_2(n) & n_2 < n \leq n_3,\\
      P_3(n) & n_3 < n,
    \end{cases}\\
    \begin{aligned}
      P_i(n) &= K_i n^{\Gamma_i}, &
      K_i &= \frac{P(n_i)}{n_i^{\Gamma_i}}.
    \end{aligned}
  \end{gather}
\end{subequations}
This introduces six new parameters: $\{n_c, n_2, n_3, \Gamma_1, \Gamma_2, \Gamma_3\}$, with the coefficients $K_i$ fixed by demanding that $P(n)$ be continuous.
For our analysis, we use the uniform priors listed in \cref{tab:eos_prior} for these.
The polytropic indices $\Gamma_i$ control the stiffness of the \gls{EoS}.
$\Gamma_i=0$ corresponds to a first-order phase transition of size $\Delta n=n_{i+1}-n_i$.
As larger polytropic indices could potentially violate causality, we switch to the causal \gls{EoS}~\cite{Lattimer:1990zz} whenever causality is violated.

\begin{table}[ht]
  \caption{\label{tab:eos_prior}%
    Priors for the core \gls{EoS} parameters (following Ref.~\cite{Raaijmakers:2019}),
    pure-neutron-matter
    parameters, and $u_p$.
    Note that the constraint $n_c \equiv n_1 < n_2$ means that the range of $n_2$ depends on
    the value of $n_c$: i.e., the range of $n_2$ differs for models \FRZ{}{1.5} and \FRZ{}{2.0}. 
    Most priors are uniform, but $a_1$ and $a_2$ follow normal distributions truncated at $\pm 2\sigma$.
    The parameter $\zeta$ (last row) is only relevant for the interpolated model \FRZ{\zeta}{}.
    }
  \begin{ruledtabular}
    \begin{tabular}{ccc}
      \textrm{Parameter} & \textrm{Range} & {\textrm{Distribution}}\\
      \colrule
      $\Gamma_1$ & [1, 4.5] & Uniform\\
      $\Gamma_2$ & [0, 8] & Uniform\\
      $\Gamma_3$ & [0.5, 8] & Uniform\\
      $n_{2}$ & [$n_c$, $8.3n_0$] & Uniform\\
      $n_{3}$ & [$n_{2}$, $8.3n_0$] & Uniform\\
      $a_1, a_2$ & [-2, 2] & Normal(0,1) \\
      $u_p$ & [2.5, 3.7] & Uniform \\
      \hline
      $\zeta$ & [0, 1] & Uniform
    \end{tabular}
  \end{ruledtabular}
\end{table}

\subsection{Parameter Space}\label{sec:eosparam}
As described at the end of the introduction, we explore models
\begin{gather*}
  \FRZ{\zeta}{n_c/n_0}
\end{gather*}
where $\zeta$ interpolates between $\FRZ{\zeta=0}{}\equiv\FRZ{\chi}{}$ and $\FRZ{\zeta=1}{}\equiv \FRZ{\phi}{}$.
For each model, we consider core-transition densities $n_c=1.5n_0$ and $n_c = 2.0n_0$ as fiducial values to explore the extent to which astrophysical observables rely on low-density nuclear inputs and vice versa.
Strong astrophysical signals may inform where low-energy nuclear models break down.

Our models are defined by more than 15 parameters (see Table I in~\cite{Forbes:2019} and Table II in~\cite{Raaijmakers:2020}).
As sampling in such a large space is computationally prohibitive, we limit the number of parameters by performing a global sensitivity analysis based on Sobol indices~\cite{saltelli2010variance} as detailed in \cref{sec:glob-sens-analys}.
Fortunately, more than half of the parameters do not significantly influence the
astrophysical observables.
This is consistent with the local sensitive analysis of~\cite{Forbes:2019}, but accounts for potential correlations among parameters.

We thus only sample the most significant parameters (beyond $n_c$) listed in \cref{tab:eos_prior} along with their priors.
These include the 5 remaining inner core parameters, the two parameters $a_{1, 2}$ describing the variations in pure neutron matter, and $u_p$ -- a parameter characterizing the self-energy of proton polarons (for detail see~\cite{Forbes:2019}).
We fix the remaining insignificant parameters -- such as those related to the symmetric nuclear matter -- at the physically motivated fiducial values described in \cite{Forbes:2019}.


\section{Bayesian Inference from Astrophysical Observations}\label{sec:BI}

To determine how well the available astrophysical data constrain our model, we use
Bayesian inference to successively update the priors listed in \cref{tab:eos_prior} with
the maximum mass constraint from the
\gls{PSR}~\mysc{j0740+6620}~\cite{2021ApJ...915L..12F}, tidal deformability
constraints from \gls{GW170817}~\cite{abbott2018gw170817}, radius measurements of \gls{PSR}~\mysc{j0030+0451}
from \gls{NICER} by \textcite{Miller:2019cac} \& \textcite{Riley:2019yda}; and radius measurements of \gls{PSR}~\mysc{j0740+6620}
from \gls{NICER} \& \gls{XMM}-Newton spectroscopy, by the same two groups \cite{Miller2021, riley2021nicer}.

Let $\Upsilon =(a_j, n_i, \Gamma_i, \dots)$ be the set of \gls{EoS} parameters listed in \cref{tab:eos_prior}, and $d$ the set of astrophysical data.
We express our prior knowledge of the parameters in terms of a \gls{PDF} $p(\Upsilon)$.
Bayes' theorem tells us how to use data $d$ to update our knowledge of the \gls{EoS} from the probability $p(d|\Upsilon)$ of obtaining the data $d$ given the parameters $\Upsilon$:
\begin{align}
  \label{eq:bayesthm}
  p(\Upsilon|d) &=
  \frac{\overbrace{p(d|\Upsilon)}^{\mathcal{L}_d(\Upsilon)}p(\Upsilon)}{Z}, &
  Z &= \int d\Upsilon\; \overbrace{p(d|\Upsilon)}^{\mathcal{L}_d(\Upsilon)}p(\Upsilon).
\end{align}
We refer to the updated knowledge $p(\Upsilon|d)$ as the posterior and $Z$ as the evidence.
It is common to call $p(d|\Upsilon) = \mathcal{L}_{d}(\Upsilon)$ a likelihood but note that the latter is not a \gls{PDF} for $\Upsilon$.

As $Z$ simply follows from normalizing the posterior, it is often left implicit, but it provides information about how well the data $d$ supports competing hypotheses (i.e., different models).
Ideally, the hypotheses correspond to disjoint regions $H_1$ and $H_2$ of parameter space, with $H_1 \cup H_2 = \varnothing$: i.e., model 1 corresponds to $\Upsilon \in H_1$ while model 2 corresponds to 
$\Upsilon \in H_2$.
This is the case, e.g., with $\FRZ{\chi}{}$ and $\FRZ{\phi}{}$, which are disjoint because of $\zeta \in \{0, 1\}$.
The degree to which the data $d$ favors one hypothesis over the other can be expressed in terms of the ratio $\mathcal{B}_{12}$ called the Bayes factor, where
\begin{gather}\label{eq:Bfact}
  \mathcal{B}_{12} = \frac{Z_{H_1}}{Z_{H_2}} \qquad
  Z_{H_{1,2}} = \int_{\Upsilon\in H_{1,2}} \!\!\!\!\!\!\!\!\! d\Upsilon\; p(d|\Upsilon)\,p(\Upsilon).
\end{gather}
If the Bayes factor is high ($\mathcal{B}_{12} \gtrapprox 3.2$), then we say that the dataset $d$ favors hypothesis $H_1$ over $H_2$, and {\it vice versa} for $\mathcal{B}_{21}\equiv 1/\mathcal{B}_{12}\gtrapprox 3.2$. For Bayes factors close to unity 
one concludes that the data $d$ does not strongly favor one hypothesis over the other~\cite{Kass:1995}.

\subsection{Heavy Pulsar \textsc{psr j0740+6620}}\label{sec:BIP}
P\mysc{SR}~\mysc{j0740+6620} is one of the heaviest known pulsars and therefore places a constraint on the maximum mass $M_{\max}(\Upsilon)$ of neutron stars supported by the \gls{EoS} characterized by $\Upsilon$. Therefore, allowed \gls{EoS}s must support neutron star masses, $M$, that are at least as heavy as $M_{\max}$. 
To impose this maximum-mass constraint on $\Upsilon$ we first construct
the \gls{PDF} of
neutron star masses, $p(M, \Upsilon|\text{Pulsar})$, by multiplying the posterior \gls{PDF} of \textsc{psr j0740+6620}, $\sim\mathcal{N}(2.08 M_{\odot}, 0.07M_{\odot} )$~\textcite{2021ApJ...915L..12F}, with the
Heaviside step function $H\bigl(M_{\text{\text{max}}}(\Upsilon)-M\bigr)$.
The likelihood function for the massive pulsar in~\cref{eq:bayesthm} is then computed as
\begin{align}
\begin{split}
    \mathcal{L}_{\text{Pulsar}}(\Upsilon) =& \int \d{M} \; p\bigl(M, \Upsilon\big|\text{Pulsar}\bigr) \; p(M) ,
\end{split}
\end{align}
where $p(M)$ is the prior on neutron star masses, which we take to be uniform on [1,3]$M_{\odot}$. 


\begin{figure*}[ht]
  \glsunset{CI}
  \centering
  \includegraphics[width=\textwidth]{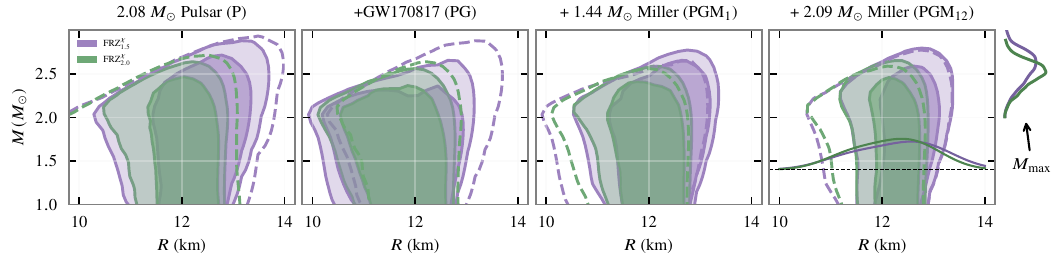}
  \caption{\label{fig:MR_DFRZ15_DFRZ20}%
    Evolution of the mass-radius posteriors for models \FRZ{\chi}{1.5} and \FRZ{\chi}{2.0} as one successively adds, from left to right, mass measurements of the \num{2.08}$M_{\odot}$ pulsar \mysc{PSR J0030+0451}~\cite{2021ApJ...915L..12F}, \gls{GW170817}, and the results in \textcite{Miller:2019cac, Miller2021} based on the 2019 (\mysc{PGM}\textsubscript{1}) and 2021 \gls{NICER} (\mysc{PGM}\textsubscript{12}) papers.
    In the left panel, the dashed lines correspond to  $2\sigma$ (\num{95}\%) priors and the nested solid lines / shaded regions correspond to the $1\sigma$ (\num{68}\%) and $2\sigma$ posteriors after including the \num{2.08}$M_{\odot}$ mass measurement.
    These posteriors become the dashed priors in the subsequent panels, and the shaded regions contain the updated $1\sigma$ and $2\sigma$ posteriors after including the subsequent observations.
    The posteriors in the right two panels correspond to posteriors \mysc{PGM}\textsubscript{1} and \mysc{PGM}\textsubscript{12} respectively in later plots.
    The rightmost panel shows posterior distributions for $R_{1.4M_{\odot}}$ and $M_{\text{max}}$.} 
\end{figure*}

\subsection{Tidal deformability from \textsc{gw170817}}\label{sec:BIG}
Analyses of the gravitational-wave signal from \gls{GW170817}~\cite{abbott2018gw170817} provided information on the tidal deformabilities $\Lambda_{1,2}$ of the two neutron stars (with primary and secondary masses $m_1$ and $m_2$, respectively) involved in that binary merger. The posteriors of these observables depend on the prior used for the magnitude of the spins of the neutron stars. Based on the current catalog of observed pulsars, we understand that the magnitude of the spin of neutron stars needs to be less than \num{0.05} if they are to merge in a Hubble time~\cite{abbott2018gw170817}. Additionally, it can be assumed that both the components of \gls{GW170817} have a common \gls{EoS}, which constrains the two $\Lambda$s. The posteriors on the masses and tidal deformabilities are then obtained through Bayesian inference assuming a specific waveform model (such as PhenomPNRT, TaylorF2, etc.)~\cite{abbott2018gw170817}.


While the imprint of the \gls{EoS} in the \gls{GW} signal of a binary neutron star is mainly sensitive to $\widetilde{\Lambda}$, which is a particular combination of $\Lambda_{1,2}$~
\footnote{This combination is just $\widetilde{\Lambda}= \frac{16}{13}\frac{(m_1 + 12 m_2)m_1^4\Lambda_1 + (m_2 + 12 m_1)m_2^4\Lambda_2}{(m_1+m_2)^5}$}, one can use the so-called \gls{EoS} insensitive relations~\cite{yagi2016binary, maselli2013equation} to disentangle the degeneracy in the influence of 
$\Lambda_{1,2}$ on 
$\widetilde{\Lambda}$
and obtain a joint probability distribution for $(m_1,m_2,\Lambda_1,\Lambda_2)$ 
from the \gls{GW170817} data. Alternatively, this distribution can be computed in terms of
$(q,\mathcal{M},\Lambda_1,\Lambda_2)$, where  $q=m_2/m_1$ is the  mass ratio and $\mathcal{M} = (m_1m_2)^{3/5}/(m_1+m_2)^{1/5}$ is the  chirp mass.
In principle, employing the \gls{EoS} insensitive relations introduces a complex dependency on the \gls{EoS} parameters $\Upsilon$ but, as their name implies, these relations depend only weakly on the \gls{EoS}.
For the mass range implied by \gls{GW170817}, one can marginalize
over \gls{EoS} models,
thereby introducing only a small fractional error of less than \num{6.5}\% each for $\Lambda_1$ and $\Lambda_2$~\cite{yagi2016binary} (see~\cite{abbott2018gw170817} for details).
The posterior $p(q, \mathcal{M}, \Lambda_1, \Lambda_2|\mysc{GW})$ was obtained in this
way~\cite{GWData} 
by using the PhenomPNRT~\cite{abbott2018gw170817} waveform,  assuming a common \gls{EoS} and without the maximum-mass constraint.
We next use this posterior to constrain $\Upsilon$.

Since the chirp mass
is well determined to be 
$\mathcal{M}_{0}=1.186(1) M_{\odot}$~\cite{TheLIGOScientific:2017qsa}, we can essentially
fix it and only need to marginalize over the poorly-constrained mass ratio $q =
m_2/m_1$ to obtain the \gls{GW} likelihood:
\begin{align}
  \mathcal{L}_{\mysc{GW}}(\Upsilon) =& \int \d{q}\d{\mathcal{M}}\;
  p(q,\mathcal{M}) \times \nonumber\\ 
  &\quad p\bigl(q, \mathcal{M}, \Lambda_1(\Upsilon, q, \mathcal{M}),
    \Lambda_2(\Upsilon, q, \mathcal{M})\big|\mysc{GW}\bigr) \nonumber\\
  \approx \int \d{q}\; p(q)\,
                                     & p\bigl(q, \mathcal{M}_0, \Lambda_1(\Upsilon, q, \mathcal{M}_0),
    \Lambda_2(\Upsilon, q, \mathcal{M}_0)\big|\mysc{GW}\bigr)
\end{align}
\exclude{To evaluate the above integral, we sample points from the priors of  $q$ and $\mathcal{M}$. These sampled points are then used to 
obtain $\Lambda_{1,2}$ of relativistic stars constructed using \gls{EoS} model with parameters $\Upsilon$.
The fact that the two masses can be expressed in terms of the well-determined chirp mass $\mathcal{M}=1.186(1) M_{\odot}$~\cite{TheLIGOScientific:2017qsa} and the mass ratio $q\equiv m_2/m_1$ reduces the above integration in the two mass parameters effectively to the one in the mass ratio.}
where $p(q)$ is a prior that we take to be uniform on [0.5, 1.0].

\begin{figure}[tb]
  \centering
  \includegraphics[width=\columnwidth]{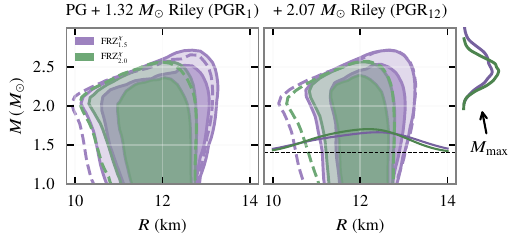}    
  \caption{%
    Same as the right two panels of \cref{fig:MR_DFRZ15_DFRZ20}, but replacing the calculations of \textcite{Miller:2019cac, Miller2021} with those of \textcite{Riley:2019yda, riley2021nicer}.
  }
  \label{fig:MR_DFRZ15_DFRZ20b}
\end{figure}

\subsection{N\textsc{icer} measurements}\label{sec:BIN}

Finally, we consider calculations based on \gls{NICER} measurements of
\gls{PSR}~\mysc{J0030+0451}~\cite{Miller:2019cac, Riley:2019yda} in 2019, and both \gls{NICER}
and \gls{XMM}-Newton data for
\gls{PSR}~\mysc{J0740+6620}~\cite{Miller2021, riley2021nicer} in 2021.
The inferred mass-radius with \num{68}\% credible intervals (\glspl{CI}) ~\cite{gardner1989confidence} are, respectively,
($\num{1.44}_{-0.14}^{+0.15}M_{\odot}$,
$\num{13.02}_{-1.06}^{+1.24}\,$\unit{km})~\cite{Miller:2019cac} and 
($\num{1.32}_{-0.18}^{+0.20}M_{\odot}$,
$\num{12.77}_{-1.45}^{+1.35}\,$\unit{km})~\cite{Riley:2019yda} for \gls{PSR}~\mysc{J0030+0451};
and ($\num{2.09}_{-0.09}^{+0.09}M_{\odot}$, $\num{13.59}_{-1.38}^{+2.33}\,$\unit{km})~\cite{Miller2021} and
($\num{2.07}_{-0.07}^{+0.07}M_{\odot}$,
$\num{12.39}_{-0.98}^{+1.29}\,$\unit{km})~\cite{riley2021nicer} for \gls{PSR}~\mysc{J0740+6620}.
These give two posteriors of the form $p\bigl(M, R|\mysc{NICER}\bigr)$ that
we use for computing the likelihood
\begin{gather}
  \mathcal{L}_{\mysc{nicer}}(\Upsilon)
   = \int \d{M}\;  p(M)\; p\bigl(M, R(\Upsilon, M)\big|\mysc{NICER}\bigr)\,,
\end{gather}
where $p(M)$ is the mass prior and is taken to be uniform on $[1, 2]M_{\odot}$ for
\gls{PSR}~\mysc{J0030+0451}, and uniform on $[1.5, 2.5]M_{\odot}$~\cite{dietrich2020multimessenger} for \gls{PSR}~\mysc{J0740+6620}. The analysis is insensitive to switching to wider priors.

We combine constraints from the maximum mass discussed in~\cref{sec:BIP}, 
\gls{GW170817} discussed in~
\cref{sec:BIG}, 
and X-ray observations in \cref{sec:BIN}
to obtain multimessenger posteriors on the \gls{EoS},
from which we compute the posteriors on the mass-radius relation shown in \cref{fig:MR_DFRZ15_DFRZ20}. 
When using 
the first \gls{NICER} measurement reported in \textcite{Miller:2019cac} 
we label our posterior as \mysc{PGM}\textsubscript{1}. Further inclusion of the second measurement ~\textcite{Miller2021} leads to posteriors labeled \mysc{PGM}\textsubscript{12}.
Following a similar prescription, the mass-radius constraints reported in \textcite{Riley:2019yda} and \textcite{riley2021nicer} are successively included in a separate analysis to obtain multimessenger posteriors labeled \mysc{PGR}\textsubscript{1} and \mysc{PGR}\textsubscript{12} shown in \cref{fig:MR_DFRZ15_DFRZ20b}. 
See~\cite{miller_2019_3473466, miller_2021_4670689, riley_2019_3386449, riley_2022_7096886} 
for details about \gls{NICER} posteriors
$p\bigl(M, R(\Upsilon, M)\big|\mysc{NICER}\bigr)$.

The Bayesian sampling in this work is performed using the PyMultiNest package~\cite{2014Buchner}, a Python implementation of the MultiNest sampling method~\cite{2009Feroz}.
When the astrophysical observations $p(d| \psi)$ are given as discrete samples over the space of astrophysical source parameters $\psi$, we use \glspl{KDE} to compute the data posterior for constructing
likelihoods. 
The bandwidth for the \gls{KDE} is chosen using Scott's rule~\cite{scott2015multivariate}.


\section{Results and Discussion}\label{sec:results}

\glsreset{CI} \Cref{fig:MR_DFRZ15_DFRZ20} and \Cref{fig:MR_DFRZ15_DFRZ20b} show $1\sigma$ and $2\sigma$ multimessenger posteriors
in the mass-radius plane.
As has been previously observed in the literature, it is evident that \gls{GW170817} favors softer
\gls{EoS}~\cite{TheLIGOScientific:2017qsa} while pulsar mass and radius measurements exclude the very soft
\gls{EoS}s~\cite{Miller:2019cac, Riley:2019yda, Miller2021,  riley2021nicer}. 
This apparent tension manifests as alternating shifts in the posteriors as the set of observations are
sequentially included from left to right in \cref{fig:MR_DFRZ15_DFRZ20}.

We have separately investigated the \gls{NICER} analyses by 
\textcite{Miller:2019cac, Miller2021} and \textcite{Riley:2019yda, riley2021nicer}.
The radius measurement reported in \textcite{Miller:2019cac, Miller2021}
provides a tighter constraint 
than that in \textcite{Riley:2019yda, riley2021nicer},
as the former rules out more \gls{EoS}s on the soft side (c.f.~\cref{tab:DrisR14}).
This difference can be largely traced back to the assumptions underlying the two groups' analyses.
For example, Miller \textit{et al.}\ used a broader prior on the radius and employed a surface emission model with three hot spots, whereas Riley \textit{et al.}\ chose a model with two hot spots.
In our study as well, we see 
in \Cref{fig:MR_DFRZ15_DFRZ20} and \Cref{fig:MR_DFRZ15_DFRZ20b} 
that the 
difference between the two calculations shows up in the mass-radius constraints with radius estimates differing by $\lesssim\SI{0.4}{km}$ on the soft side.

\begin{table}[t]
  \newcommand{\dat}[2]{[\num{#1}, \num{#2}] \si{km}}
  \renewcommand{\dat}[2]{[\num{#1}, \num{#2}]}
  
  \def\lo{7}              
  \def\hi{13.1}           
  \def\bcolor{black!25}   
  \def\fcolor{black}      

  \newlength{\twidth}\setlength{\twidth}{\widthof{\dat{10.89}{12.94}}}  
  \newlength{\tdepth}\setlength{\tdepth}{1ex}                           
  \newlength{\pheight}\setlength{\pheight}{0.2\tdepth}                  
  \newlength{\rowheight}\setlength{\rowheight}{0.3\normalbaselineskip}  
  \newlength{\tsep}\setlength{\tsep}{0.1em}                             

  \newlength{\pwidth}\setlength{\pwidth}{\twidth + 2\tsep}              

  \newlength{\lwidth}
  \newlength{\rwidth}
  \newcommand{\mybox}[2]{%
    \pgfmathsetlength{\lwidth}{(#1 - \lo)*\twidth / (\hi - \lo)}%
    \pgfmathsetlength{\rwidth}{(#2 - #1)*\twidth / (\hi - \lo)}%
    \rlap{\textcolor{\bcolor}{\rule[-\tdepth]{\pwidth}{\pheight}}}%
    \hspace{\lwidth}%
    \textcolor{\fcolor}{\rule[-\tdepth]{\rwidth}{\pheight}}%
  }
  \WithSuffix\newcommand\dat*[2]{\strut\rlap{\mybox{#1}{#2}}\hspace{\tsep}\dat{#1}{#2}}
  \newcommand{\mylabel}[2][]{$\qquad\llap{#1} + {}$#2}
  \caption{\label{tab:DrisR14}%
    \textbf{Inferred Radius of a $\mathbf{1.4M_{\odot}}$ Neutron Star:}
    \num{95}\% \gls{CI} on the radius (in \unit{km}) of a $\num{1.4}M_{\odot}$ neutron star
    for \FRZ{\chi}{1.5} and \FRZ{\chi}{2.0} as one progressively includes astrophysical
    observations.
    Under each range is a visual indication of the \gls{CI} (black) in the band
    \dat{\lo}{\hi} (light gray) to facilitate comparison.
    The tighter constraints using the analysis of \textcite{Miller:2019cac, Miller2021}
    are highlighted in gray:
    $R_{1.4}=\num{12.46}^{+0.48}_{-1.02}\,$\unit{km} for $n_c=\SI{1.5}{n_0}$ and $12.28_{-0.86}^{0.50}$ \unit{km} for $n_c=\SI{2.0}{n_0}$ and shown in \cref{fig:R14}.
  }

    \begin{tabular}{lp{\pwidth}p{\pwidth}}
      \toprule
      
      &\multicolumn{2}{c}{%
        $R_{1.4M_{\odot}}$ for \FRZ{\chi}{n_c}
        {(\num{95}\% \gls{CI} [\unit{km}])}}\\
      Source & $n_c=\SI{1.5}{n_0}$ & $n_c=\SI{2.0}{n_0}$\\
      \midrule
      Prior                                 & \dat*{7.85}{13.05}  & \dat*{8.43}{12.79}\\[\rowheight]
      \mylabel{$2.08 M_{\odot}$ (\textsc{j0030+0451})}     & \dat*{10.80}{13.02} & \dat*{10.81}{12.73}\\[\rowheight]
      \mylabel{\gls{GW170817} (= PG)}       & \dat*{10.27}{12.82} & \dat*{10.50}{12.65}\\[\rowheight]
      \textbf{Miller \textit{et al.}} & {} & {}\\[\rowheight]
      \mylabel[PG]{$1.44 M_{\odot}$ (\textsc{j0740+6620})} & \dat*{10.89}{12.94} &
                                                                    \dat*{10.91}{12.72}\\[\rowheight]
      \cellcolor{lightgray}
      \mylabel{$2.09 M_{\odot}$ (\textsc{j0030+0451})}     & \dat*{11.44}{12.95} & \dat*{11.42}{12.78}\\[\rowheight]
      \textbf{Riley \textit{et al.}} & {} & {}\\[\rowheight]
      \mylabel[PG]{$1.32 M_{\odot}$ (\textsc{j0740+6620})} & \dat*{10.55}{12.88} & \dat*{10.74}{12.69}\\[\rowheight]
      \mylabel{$2.07 M_{\odot}$ (\textsc{j0030+0451})}     & \dat*{11.05}{12.90} & \dat*{11.13}{12.72}\\[\rowheight]
      \bottomrule
    \end{tabular}
\end{table} 

\begin{figure}[t]
  \centering
  \includegraphics[width=\columnwidth]{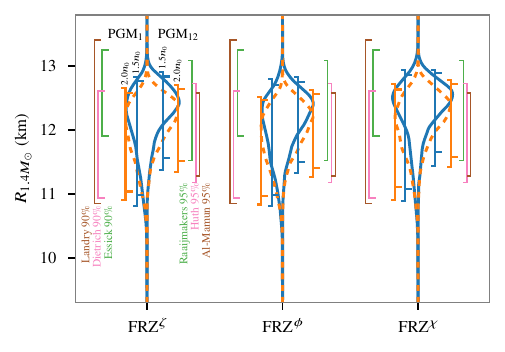}
  \caption{\label{fig:R14}%
    A pair of violins is plotted above for each of the  \FRZ{\zeta}{}, \FRZ{\phi}{}, and \FRZ{\chi}{} models. Each pair consists of two violins -- the blue (solid) one is for the transition density of 1.5 $n_0$ and the orange (dashed) one is for the transition density of 2 $n_0$.
    The left contour of each violin plots the constraint on the radius of a 1.4
    $M_{\odot}$ neutron star using three astrophysical observations, namely
    \num{2.08}$M_{\odot}$ pulsar (P), \gls{GW170817} (G), \gls{NICER} measurements by
    \textcite{Miller:2019cac} ($M_{1}$). On the other hand, the right
    contour plots the radius constraint after including the second \gls{NICER} measurement \textcite{Miller2021} ($M_{2}$). Double-sided brackets are overlaid for each contour and denote \num{90}\% (right facing) and \num{95}\% (left facing) \glspl{CI}. Also drawn are the single sided brackets denoting the \num{95}\% \glspl{CI} for $R_{1.4M_\odot}$ in \textcite{al2021combining}, \textcite{huth2022constraining}, \textcite{Raaijmakers:2021uju} and \num{90}\% \gls{CI} obtained by \textcite{Essick:2020flb}, \textcite{dietrich2020multimessenger}, \textcite{landry2020nonparametric}.}
\end{figure}

\begin{figure}[t]
   \centering
   \includegraphics[width=\columnwidth]{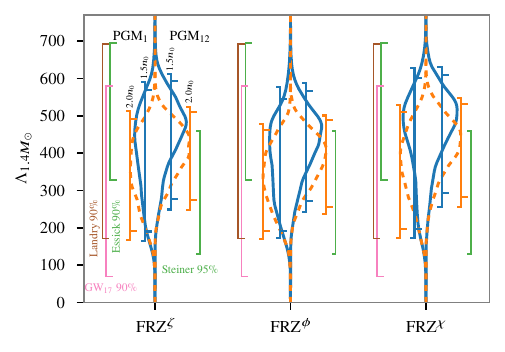}
   \caption{\label{fig:L14}%
 Same as Figure ~\ref{fig:R14} but for tidal deformability of $1.4M_\odot$ stars. Also shown are the for \num{90}\% \glspl{CI} for $\Lambda_{1.4M_{\odot}}$ from \gls{GW170817} alone~\cite{abbott2018gw170817}. Additionally, \num{90}\% \glspl{CI} from \textcite{Essick:2020flb}, \textcite{landry2020nonparametric} and \num{95}\% \gls{CI} by \textcite{PhysRevC.91.015804} are also drawn.}
\end{figure}

\Cref{tab:DrisR14} shows the \num{95}\% \glspl{CI} of the radius of a $1.4
M_{\odot}$ neutron star for the models \FRZ{\chi}{2.0} and \FRZ{\chi}{1.5}.
These results are consistent with existing constraints in the literature.
Comparisons of multimessenger posteriors on $R_{1.4M_\odot}$ and $\Lambda_{1.4M_\odot}$ are shown in \cref{fig:R14} and \cref{fig:L14} respectively. There, we see that the \num{90}\% \glspl{CI} are within \num{10}\% despite differences in the modeling of neutron-star \glspl{EoS} and the sets of astrophysical data considered.

Among the various approaches to incorporate low-density nuclear physics inputs, 
a majority of the works (e.g.~\cite{Tews2019EPJA, biswas2021impact}) assumed a quadratic expansion in isospin asymmetry that is constrained by the symmetry energy $S$ and its slope $L$. 
The recent neutron skin measurement {PREX-II} favors large $S$ and $L$, resulting in stiffer nuclear \gls{EoS} in \cite{biswas2021impact} compared to ours.

Choices for the high-density phase are abundant and 
range from the piecewise polytrope employed in~\cite{raaijmakers2021constraints}, to the Gaussian-Process-based parameterization in~\cite{landry2020nonparametric}. 
As expected, this leads to
large variations in the posterior. 
Nevertheless, as the comparisons show, the resulting \glspl{CI} on $R_{1.4M_\odot}$ and $\Lambda_{1.4M_\odot}$ converge at the level of $\sim$\num{10}\%.

The analyses cited above also considered different sets of astrophysical observations. 
\textcite{raaijmakers2021constraints} employed calculations from Miller \textit{et al.}\ whereas  \textcite{huth2022constraining} combined the constraints from Riley \textit{et al.}\ as well.
As noted earlier, this only has a minor impact (as can be seen from Figure~\ref{fig:R14}).
Additional neutron-star radii measurements from  \gls{qlmxb} \cite{heinke2006hydrogen} and \gls{pre} \cite{ozel2009mass} was incorporated in \textcite{PhysRevC.91.015804}.
These measurements may suffer from unknown systematics, and were not included in our analysis: they appear to only have minor impacts.
Lastly, the putative upper limit on neutron-star maximum mass is imposed.
Our analysis does not make use of this putative information, resulting in posterior \glspl{CI} on maximum mass considerably higher than those reported in the literature.
Despite this, the posteriors on $R_{1.4M_\odot}$ and $\Lambda_{1.4M_\odot}$ largely agree.
This suggests that the properties of canonical $1.4~M_\odot$ neutron stars largely decouple from those of the maximum mass stars even with a simple piecewise polytropic inner core \gls{EoS}. 
Indeed, the tighter bounds on $M_\mathrm{max}$ reported in the literature appear to stem from direct constraints on $M_\mathrm{max}$ itself via hypotheses based on
the post-merger evolution of \gls{GW170817}~\cite{Margalit:2017dij, Rezzolla_2018}.

\begin{table*}[ht]
  \caption{\label{tab:Evidence} Global log-evidence (to the base e) from importance sampling~\cite{Feroz:2013hea} for 6 \gls{EoS} models as 
    one progressively includes more astrophysical observations. The numbers in the brackets are sampling errors written in the scientific notation. The Bayes factors (obtained by taking the exponential of the difference between any two log evidences) for any given observation (row) are not significantly different
    and do not reveal a preference for any \gls{EoS} model.}
  \begin{ruledtabular}
    \begin{tabular}{ccccccc}
       &\multicolumn{2}{c}{\FRZ{\chi}{n_c}}&\multicolumn{2}{c}{\FRZ{\zeta}{n_c}}&\multicolumn{2}{c}{\FRZ{\phi}{n_c}}\\
       Source&$n_c$=1.5&$n_c$=2.0&$n_c$=1.5
      &$n_c$=2.0&$n_c$=1.5
      &$n_c$=2.0\\ \hline
       $2.08 M_{\odot}$ Pulsar&-2.34(6)&-2.35(6)&-2.67(7)&-2.67(8)&-2.41(6)&-2.48(4) \\
       +\gls{GW170817}&-9.07(6)&-8.69(18)&-9.00(8)&-9.07(8)&-8.91(4)&-8.77(6)\\
       \textbf{Miller \textit{et al.}} & {} & {}\\
       PG + $1.44 M_{\odot}$ Pulsar &-10.73(6)&-10.57(7)&-10.91(5)
       &-10.87(8)&-10.83(4)&-10.69(8)\\
       + $2.09 M_{\odot}$ Pulsar &-12.12(20)&-12.44(8)&-12.61(8)
       &-12.62(6)&-12.00(5)&-12.39(12)\\
       \textbf{Riley \textit{et al.}} & {} & {}\\
       PG + $1.32 M_{\odot}$ Pulsar&-10.30(15)&-10.40(5)&-10.75(6)
       &-10.61(7)&-10.45(9)&-10.24(8)\\
       + $2.07 M_{\odot}$ Pulsar&-11.93(7)&-11.68(20)&-11.82(15)
       &-11.80(20)&-11.67(14)&-11.84(7)
    \end{tabular}
  \end{ruledtabular}
\end{table*}

In this work, we propose a unified parameterization for both chiral \gls{EFT}-based and phenomenological potential-based \gls{PNM} \glspl{EoS} that allows for natural interpolations between the two nuclear models.
This framework enables model selection between the nuclear models, but the current data does not yet provide significant discrimination.
For the interpolated model \FRZ{\zeta}{}, the posterior on $\zeta$ marginally shifts towards the \gls{EFT}-based description, indicating a slight preference for it.
This can be largely traced back to the stiffer prior the \gls{EFT} model admits compared to that of the phenomenological potential-based model (see \cref{fig:zeta}).
We also calculated the log-evidence for \FRZ{\chi}{}, \FRZ{\phi}{} and \FRZ{\zeta}{}, which are listed in \cref{tab:Evidence}.
As mentioned above, the Bayes factor \cref{eq:Bfact}
can facilitate model selection if it deviates noticeably from unity, by at least a factor of $\sim3$ (or $1/3$).
The log-evidence listed in \cref{tab:Evidence}
do not reveal a significant preference for any of the models suggesting a limited discerning power by current astrophysical observations.
We defer to future work on the required error budget for model selection.

\begin{figure}
   \centering
   \includegraphics{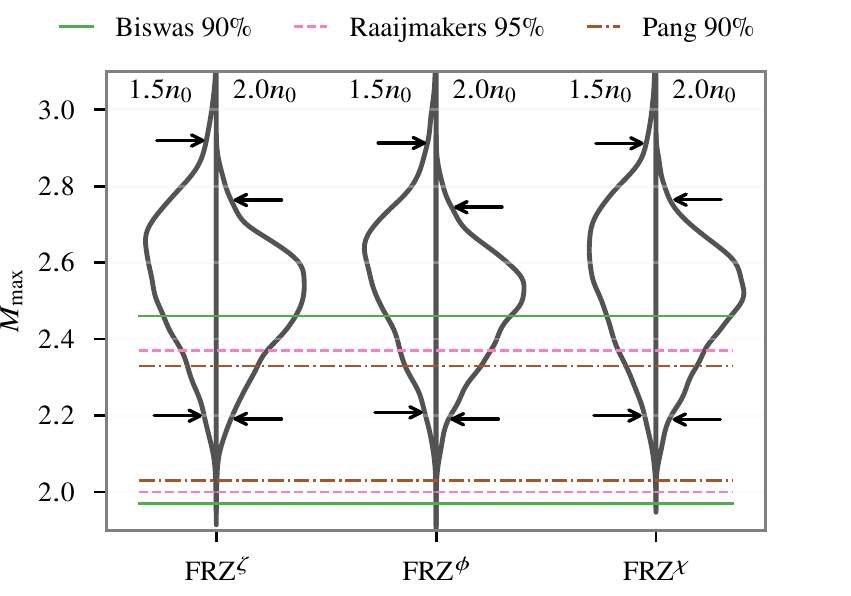}
    \caption{Same as Figure ~\ref{fig:R14} but for $M_{\mathrm{max}}$ of neutron stars. Also shown are the \num{90}\% \gls{CI} for $M_{\mathrm{max}}$ by \textcite{Essick:2020flb}, \textcite{landry2020nonparametric} and \num{95}\% \gls{CI} by \textcite{PhysRevC.91.015804}.}
    \label{fig:M14}
\end{figure}

\begin{acknowledgments}
We thank S. Gandolfi for providing an extended set of QMC phenomenological potential-based \glspl{EoS}. We also thank Tim Dietrich for carefully reading the manuscript and sharing useful comments.
We gratefully acknowledge the use of the high-performance super-computing clusters Kamiak and Pegasus at Washington State University (WSU) and The Inter-University Centre for Astronomy and Astrophysics (IUCAA), respectively. B.B. acknowledges the support from University Grants Commission (UGC), India, the Knut, and Alice Wallenberg Foundation 
under grant Dnr.~KAW~\textlf{2019.0112}, and the Deutsche 
Forschungsgemeinschaft (DFG, German Research Foundation) under 
Germany's Excellence Strategy – EXC \textlf{2121} ``Quantum Universe'' –
\textlf{390833306}.
This material is based upon work supported by \gls{NSF}'s \gls{LIGO} Laboratory, which is a major
facility fully funded by the \gls{NSF}.
P.T., M.M.F. and S.B. acknowledge support from the \gls{NSF} under Grants \mysc{PHY-2012190} and
\mysc{PHY-2309352}.
\end{acknowledgments}

\appendix
\newpage
\section*{Appendices}
\label{sec:appendix}

\begin{table*}[ht]
\caption{\label{tab:poly_coef}%
Coefficients of the polynomials fitted to \dEFT{} data \cite{Drischler:2020} and phenomenological data \cite{Gandolfi:2012} along with the polynomial fit to the two most dominant eigenvectors of the covariance matrix of the residual energy (as described in \cref{sec:QMCData}).
}
\begin{ruledtabular}
\begin{tabular}{cccccc}
\textrm{Polynomial}& {} & {} &
\textrm{Coefficients}\\
{} & $c_4$ & $c_3$ & $c_2$ & $c_1$ & $c_0$ \\
\colrule
\\
$\Eres_{0}^{\phi}$ & -172.33 &  660.87 & -905.05 &  547.612 &
        -140.25\\
$\Eres_{0}^{\chi} $ & 7.8261 & -22.787 & 44.778 & -30.786 &  -7.5286 \\
$\Eres_1^{\chi}$ & -2.1849 &  14.750 & -22.012 & 13.253 &
          -2.8860 \\
$\Eres_2^{\chi}$ & 11.356 & -30.849 & 30.301 & -13.428 &
           2.2623
\end{tabular}
\end{ruledtabular}
\end{table*}

\subsection{Basis from \gls*{chiEFT} Data}\label{sec:ChiData}

\begin{figure}
      \centering
      \includegraphics[scale=0.9]{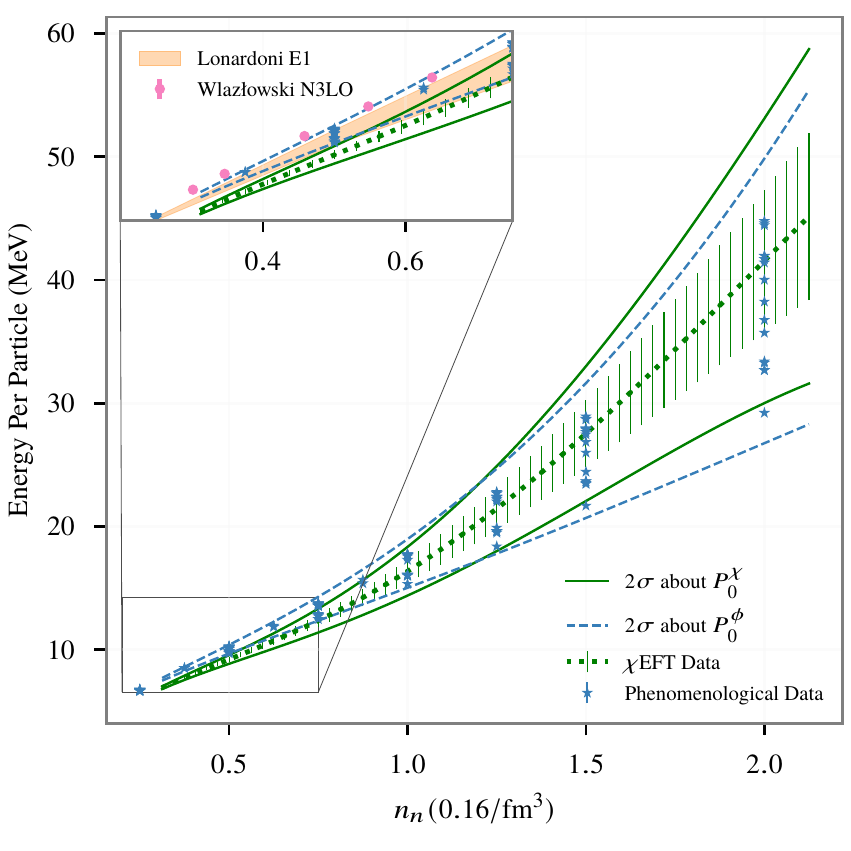}
      \caption{Comparing the \dEFT{} data \cite{Drischler:2020} (green dotted line
        with vertical error bars) and phenomenological data \cite{unpublishedGandolfi}
        (blue stars) with the fits obtained from polynomial models for pure neutron
        matter.  Solid green (dashed blue) line shows the 2 $\sigma$ energy bounds about
        mean energy of \dEFT{} data (phenomenological data). Inset plot points out the
        inconsistencies among different calculations in the energy values of pure
        neutron matter in the low-density region.
      }
      \label{fig:Eig_Func}
  \end{figure}

\textcite{Drischler:2020} used the Gaussian process to fit the data from many-body perturbation theory calculations of \gls{chiEFT}~\cite{PhysRevLett.122.042501} at various orders of expansion in the ratio of momentum and breakdown scale. Their fitting was performed to account for the truncation errors at increasing orders of expansion. In this work, a breakdown scale of $600$ MeV and a momentum cutoff of $500$ MeV were chosen to obtain the correlated energy values at selected density points (see \cref{fig:Eig_Func}). Now, since the energy is parameterized as
\begin{align}
  E_N = m_{n}c^2 + \kappa \bar{k}^2 + \bar{k}^3 \Eres(\bar{k}),
  \label{Eq:EPNM2}
\end{align}
and our aim is to model $\Eres$, we calculate the covariance matrix, $\Sigma$, for $\Eres$ from correlated energy values. 

We found that the largest two eigenvalues of $\Sigma$ -- termed $d_1$ and $d_2$ -- are much larger than the remaining ones. Qualitatively, this means that the variations 
in $\Eres$
 are significantly stronger along their corresponding eigenvectors than the rest.
 This observation enables us to represent the variation in $\Eres$ as
 \begin{equation}
     \delta \Eres = a_1\sqrt{d_1} V_1 + a_2\sqrt{d_2} V_2.
 \end{equation}
 Here, $V_i$ are the two dominant eigenvectors of $\Sigma$, and $a_i$ are distributed as $\mathcal{N}(0, 1)$. Therefore, $\Eres$ modeled through \dEFT{} data is given by
 \begin{equation}
     \Eres^{\chi} = \Eres_{0}^{\chi} + \delta \Eres,
 \end{equation}
 where $\Eres_{0}^{\chi}$ is obtained by transforming the mean energy of \gls{chiEFT} data. We fit $\sqrt{d_{1,2}} V_{1,2}$ and $\Eres_{0}^{\chi}$ with polynomials
 in $\bar{k}$.
 The order of these polynomials (namely, $\Eres_{1}^{\chi}, \Eres_{2}^{\chi}$ and $\Eres_{0}^{\chi}$) is found to be 4, which is the minimum order that satisfies 
 our requirement that the maximum variance in the error for fitting to the mean energy and 
 eigenvectors is less than $10^{-4}$.

 \subsection{Basis from Phenomenological Potential Based Data}\label{sec:QMCData}

 The phenomenological energy data is discrete and therefore the double derivatives cannot be calculated precisely. This makes the corresponding speed of sound arbitrary. The densities at which energy is provided also vary as we go from one potential to the other. Therefore, we create bins of densities corresponding to the union of density points at which the energies are given. Then, we fit a fourth-order polynomial to individual $\Eres$ data to find the missing $\Eres$ in these bins. Finally, we find the mean of $\Eres$ in all these bins and fit a fourth-order polynomial, in $\bar{k}$, to this mean energy (call it $\Eres_0^{\phi}$).

 We find that the variations in the phenomenological $\Eres$ can be captured using the eigenvectors of the covariance matrix $\Sigma$. That gives the representation of $\Eres$ for phenomenological data as
 \begin{equation}
     \Eres^\phi = \Eres^\phi_0 + \sum_{k=1}^2 c_k \Eres^\chi_k.
 \end{equation}
 Here, $c_1$ and $c_2$ are of order 1. We provide the coefficients of the polynomial fits to the four $\Eres$ in Table \ref{tab:poly_coef}.

 The models for the phenomenological and \gls{chiEFT} data can be bridged through a translation parameter, $\zeta$. More precisely,
 \begin{align}
   \label{eq:polytrope}
   \Eres &= \sum_{k=1}^2 a_k \Eres^\chi_k + \Eres^\chi_0 + \zeta (\Eres^\phi_0 - \Eres^\chi_0).
\end{align}
So, our model has three parameters, $a_1, a_2$, and $\zeta$. Note that $\zeta=0$ gives $ \Eres(\bar{k}) = \Eres^{\chi}$.
 Using $\Eres(\bar{k})$ leads to employing an additional parameter during our Bayesian inference.

 One of the main advantages of the model presented here, which is absent in other models based on simulation data, is that it 
 conserves the speed of sound information from the data. And the procedure is general enough to apply to any data in the future. This also provides a natural way of comparing data among different simulations, in light of various microscopic or macroscopic observations.

  \begin{figure}
      \centering
      \includegraphics[scale=0.9]{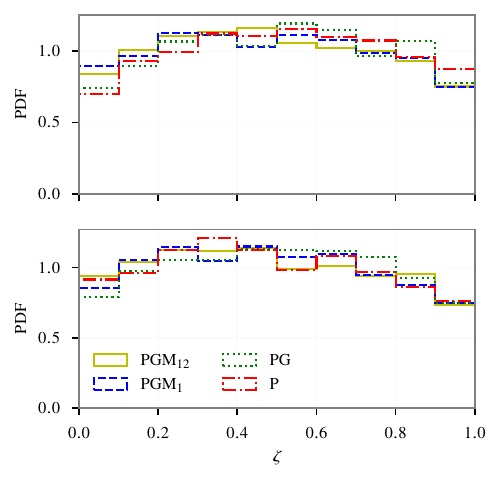}
      \caption{Comparing the $\zeta$ distribution after successively adding astrophysical observations for the \FRZ{}{\zeta} models for transition to core at $1.5 n_0$ (top) and $2 n_0$ (bottom) . The posteriors are dominated by prior and that is why it is flat. This is consistent with the fact that the Bayes' factor between \FRZ{\chi}{1.5} (i.e. $\zeta$=0) and \FRZ{\phi}{1.5} (i.e. $\zeta$=1) is $\sim$ 1 i.e both the extreme points are equally favored. The decrease in the probability at the extreme values are edge effect that can be treated by reweighing the posterior samples with prior samples. 
      }
      \label{fig:zeta}
  \end{figure}

\begin{figure*}[tb]
    \begin{minipage}[t]{\columnwidth}
    \centering
        \includegraphics[width=\columnwidth]{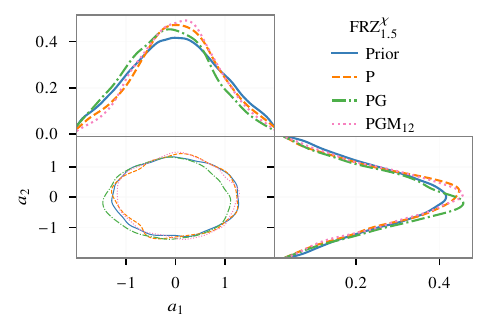}
    \end{minipage}
    \hfill
    \begin{minipage}[t]{\columnwidth}
        \includegraphics[width=\columnwidth]{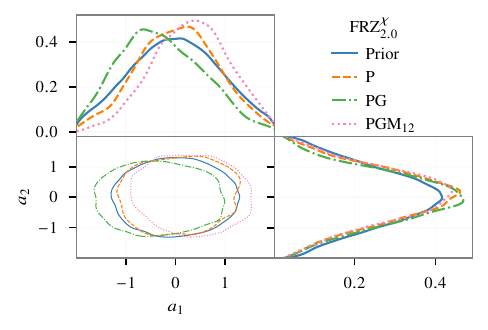}
    \end{minipage}
    \caption{The posteriors of pure-neutron-matter parameters when progressively including  the pulsar (P)~\cite{Cromartie:2019kug} followed by \gls{GW170817} (PG)~\cite{TheLIGOScientific:2017qsa}, and finally the multiple \gls{NICER} (\mysc{PGM}\textsubscript{12})~\cite{Miller:2019cac, Miller2021} observations for the \FRZ{\chi}{1.5} and \FRZ{\chi}{2.0} models.}
    \label{fig:a1a2_DFRZ20}
\end{figure*}

\subsection{Bayesian Information Criteria}
As discussed in the conclusion, the available data does not have enough power to discriminate between the different nuclear models.
To see this, we plot the posterior distribution for $\zeta$ in \cref{fig:zeta} and find that it remains flat.

Although not really needed here, to formally compare the models -- including ``Occam's penalty'' that should be 
imposed on \FRZ{\zeta}{} 
for using an additional parameter -- we compute the \gls{BIC}~\cite{AKAIKE1983165}, given by, 
\begin{align}
    \mathrm{BIC} = n_{\text{par}}\ln(n_{\text{sam}}) - 2\ln(\hat{L}), 
\end{align}
where $n_{\text{par}}$ denotes the number of model parameters over which inference is being performed, $n_{\text{sam}}$ is the number of posterior samples, and $\hat{L}$ is the maximum likelihood value computed through model parameters. 
The first term signifies Occam's penalty whereas the second term quantifies how well the model fits the data. A higher \gls{BIC} value means the model is less preferred. 

\Cref{tab:BIC} gives the \gls{BIC} values for \FRZ{}{1.5} models using likelihoods defined in section~\ref{sec:BI}.
It shows that the data has no significant preference for either of the two nuclear models, $\FRZ{\chi}{1.5}$ and $\FRZ{\phi}{1.5}$, while the \gls{BIC} value for the \FRZ{\zeta}{1.5} model is consistently given by $\ln n_{\text{samp}} \approx 10$ -- demonstrating Occam's penalty.

\begin{table}[ht]  \caption{\label{tab:BIC}\glsreset{BIC}\Gls{BIC}~\cite{AKAIKE1983165} from sampling for 3 \gls{EoS} models. In each column, various combinations of astrophysical observations are considered in the different rows. A higher value indicates a lesser preference for the model. We see that \FRZ{\zeta}{1.5} gets consistently penalized for having an extra parameter.
  }
  
  \begin{ruledtabular}
    \begin{tabular}{cccc}
       Source&\FRZ{\chi}{1.5}
       &\FRZ{\zeta}{1.5}
       &\FRZ{\phi}{1.5}\\[0.3em]  
       \hline
       $2.08 M_{\odot}$ Pulsar&75.83& 84.06& 75.88 \\
       +\gls{GW170817}&97.57&105.69 &96.97\\
       \textbf{Miller \textit{et al.}} & {} & {}\\
       PG + $1.44 M_{\odot}$ Pulsar &107.00&115.25 &106.88\\
       + $2.09 M_{\odot}$ Pulsar &126.37& 135.46& 126.28\\
       \textbf{Riley \textit{et al.}} & {} & {}\\
       PG + $1.32 M_{\odot}$ Pulsar&116.46& 125.53& 116.16\\
       + $2.07 M_{\odot}$ Pulsar&125.85& 135.14& 125.78
    \end{tabular}
  \end{ruledtabular}
\end{table}

\subsection{Global Sensitivity Analysis}\label{sec:glob-sens-analys}
To obtain a Bayesian inference on the \gls{EoS} models considered in this paper, one must sample a 16- or 17-dimensional parameter space.
As more detections are included, this becomes computationally prohibitive, so we perform a global sensitivity analysis to determine the parameters that are most sensitive to the observables.
We calculate total effect index $S_T\in[0,1]$~\cite{saltelli2010variance} of each
parameter $X_i$, taking into account its prior:
\begin{align}
    S_T = \frac{E_{X_{\sim i}}\Bigl(V_{X_{i}}(Y|X_{\sim i})\Bigr)}{V(Y)}.\\
    S_T = 1 - \frac{V_{X_{\sim i}}\Bigl(E_{X_{i}}(Y|X_{\sim i})\Bigr)}{V(Y)}.
\end{align}
Here $E_f(\cdot)$ and $V_f(\cdot)$ are respectively the expectation and variance
taken over the ``factor'' $f$, which is either the parameter for which the coefficient is being
calculated ($X_i$), or all the parameters \emph{except} the current parameter ($X_{\sim i}$). 

A large $S_T$ (close to 1) means that the parameter is more sensitive to the specified
observable.
Based on this measure, we choose highly sensitive parameters for sampling in our inference and fix the less sensitive parameters to a central value described in \cite{Forbes:2019}. We find that the inner core parameters (polytropic indices and transition densities), outer core parameters ($a_1$ and $a_2$), and $u_p$ have a total effect index (w.r.t maximum mass, radius, and tidal deformability) that is at least 2 orders of magnitude higher than the remaining parameters. 


\renewcommand{\rmdefault}{ntxtlf}
\fontfamily\familydefault\selectfont
\bibliography{master_loc, mybiblio,local}

\end{document}